\documentclass[14p]{article}

\usepackage[T1]{fontenc}
\usepackage[utf8]{inputenc}
\usepackage{geometry}
\usepackage{color}
\usepackage{float}
\usepackage{amsmath}
\usepackage{graphicx}
\usepackage{multicol}
\usepackage{babel}
\usepackage{graphicx}
\usepackage{subcaption}
\usepackage{booktabs}
\usepackage{setspace}
\usepackage{url}
\graphicspath{{figures/}} 

\date{} 

\title{How null-model constraints affect statistical validation in projected bipartite networks}
\author{
Alessandro Catalano\textsuperscript{1}, Rosario N. Mantegna\textsuperscript{2,3}\\
\small 1 Department of Physics and Astronomy "Ettore Majorana", University of Catania, Italy. \\
\small 2 Department of Physics and Chemistry "Emilio Segrè", University of Palermo, Italy. \\
\small 3 Complexity Science Hub Vienna, Vienna, Austria. \\
\small Corresponding authors: alessandro.catalano@phd.unict.it, rosario.mantegna@unipa.it
}

\begin{document}

\maketitle

\begin{abstract}
Statistical validation of projected bipartite networks depends critically on the null model adopted to describe random co-occurrences. Although several null models have been proposed, their comparison has mainly focused on the validated backbones they produce rather than on the statistical assumptions underlying their construction. Here we compare four widely used null models—the microcanonical configuration model generated by the Curveball algorithm, the Bipartite Configuration Model (BiCM), the Bipartite Partial Configuration Model (BiPCM), and the Hypergeometric approximation—using three empirical bipartite systems from comparative genomics, international trade and food science. 
We use the statistically validated links obtained from the microcanonical bipartite configuration model as the reference benchmark for assessing performances of the other three models. Our central result is that the statistical consequences of relaxing null-model constraints cannot be understood solely from the constraints themselves but must be analyzed through the probability distribution induced for the co-occurrence statistic. In particular, the combined behaviour of the expectation and variance largely explains the observed differences among the statistically validated backbones. We further derive a leading-order sparse approximation for the BiCM expectation, showing that the first correction to the Hypergeometric prediction is controlled by the degree heterogeneity of the non-projected layer. Surprisingly, despite neglecting this heterogeneity, the Hypergeometric model accurately reproduces the co-occurrence variance of the microcanonical ensemble across all datasets. Our results suggest that null models should be compared not only according to the constraints they preserve but also according to the statistical consequences that these constraints induce on the distribution of the test statistic.
\end{abstract}

\section{Introduction}

Statistical validation plays a central role in the analysis of complex systems \cite{Maslov2002,Milo2002,Garlaschelli2008,Serrano2009,Lancichinetti2010,Tumminello2011,Squartini2011,Tumminello2012,Strona2014,Neal2014,Li2014,Hatzopoulos2015,Curme2015,Saracco2015,Saracco2017,Strona2018,Challet2018,Marcaccioli2019,Han2019,Cimini2019,Vallarano2021,Musciotto2021a,Musciotto2021b,Cimini2022,Neal2024} . Whenever an empirical observation is compared against random expectations, the final inference depends not only on the observed value of the statistic under investigation but also on the null hypothesis adopted to describe random behaviour. The null model determines the probability distribution against which the observation is evaluated and therefore directly affects the statistical significance assigned to the observed event. Consequently, the choice of the null hypothesis is not merely a technical aspect of hypothesis testing but an integral part of the inference procedure itself.

Projected bipartite networks provide a particularly suitable framework in which to investigate this problem. Bipartite representations naturally arise in numerous biological \cite{Tumminello2011}, ecological \cite{Strona2014,Strona2018}, economic \cite{Tumminello2011,Tumminello2012,Hatzopoulos2015,Curme2015,Saracco2015,Saracco2017,Cimini2022}, social \cite{Tumminello2012,Neal2014,Li2014,Han2019,Musciotto2021a} and socio-technical \cite{Musciotto2021b} systems, including organisms and genes, countries and traded products, recipes and ingredients, authors and scientific papers, or users and online contents. In many applications the primary interest concerns the relationships among nodes belonging to one layer of the bipartite network. These relationships are inferred through network projection, where two nodes become connected whenever they share one or more neighbours in the opposite layer. Since highly connected nodes naturally generate many co-occurrences, projected networks typically contain a large number of links that are compatible with random expectations. Statistical validation is therefore required to distinguish genuine associations from those that can be explained solely by the structural properties of the original bipartite system.

Over the last fifteen years, several approaches have been proposed to perform the statistical validation of projected bipartite networks. These approaches range from exact microcanonical randomization procedures \cite{Strona2014} to maximum-entropy canonical models \cite{Garlaschelli2008,Saracco2015,Saracco2017} and analytical approximations based on urn formulations \cite{Tumminello2011}. Although these methods differ substantially in both computational cost and the structural constraints that they preserve, they all aim at estimating the probability distribution of the co-occurrence random variable associated with pairs of nodes in the projected layer. This distribution provides the reference against which the observed overlap is evaluated and therefore determines the outcome of the statistical test.

Previous studies have primarily compared these methods by analysing the statistically validated backbones they produce \cite{Neal2014} or by introducing meta-validation procedures \cite{Cimini2019} aimed at assessing their mutual consistency. Such comparisons have provided valuable practical guidance and have clarified the robustness of projected networks obtained under different validation strategies. However, comparatively less attention has been devoted to understanding why different null models produce different statistically validated backbones. In particular, the relationship between the structural constraints defining the null model and the statistical properties of the resulting co-occurrence distribution remains only partially understood.

The present work adopts this perspective. Rather than asking which statistical validation procedure performs best, we investigate how the explicit and implicit constraints encoded in different null models propagate into the probability distribution of the co-occurrence random variable and ultimately affect statistical validation. Our analysis considers four null models representing different levels of structural description: the microcanonical configuration model implemented through the Curveball algorithm \cite{Strona2014}, the Bipartite Configuration Model (BiCM) \cite{Saracco2015}, the Bipartite Partial Configuration Model (BiPCM) \cite{Saracco2017}, and the Hypergeometric approximation \cite{Tumminello2011}. Throughout this work we use the microcanonical model as the reference benchmark because, among the considered models, it preserves exactly the observed degree sequence of both layers of the bipartite network.

A central observation emerging from our analysis is that comparing null models solely according to the constraints they preserve provides only a partial understanding of their statistical behaviour. What ultimately determines the outcome of statistical validation is the probability distribution of the co-occurrence random variable generated by each null model. Although obtaining this distribution analytically is generally difficult, and in some cases computationally prohibitive, we show that much of the observed behaviour can already be understood by analysing its first two moments. The expectation determines the location of the null distribution, whereas the variance controls its fluctuations. Their combined effect largely explains the different numbers of true positives, false positives and false negatives obtained with the different validation procedures.

The paper makes three main contributions. First, we provide a systematic comparison of four widely used null models across three bipartite systems from biology, economics and food science, thereby demonstrating that the effects of relaxing the constraints defining the null model depend strongly on the structural organization of the underlying network. Second, we derive a leading-order sparse approximation for the expected co-occurrence under the Bipartite Configuration Model. Although this approximation is quantitatively accurate only in sufficiently sparse bipartite networks, it identifies the normalized degree heterogeneity of the non-projected layer as the leading correction to the Hypergeometric expectation, thereby providing a transparent interpretation of the role played by degree heterogeneity in the statistical validation problem. Third, we show that the comparison of the first two moments of the co-occurrence distribution provides a natural framework for understanding the statistical performance of different null models and for identifying the conditions under which computationally inexpensive analytical approximations reproduce the microcanonical benchmark.

Beyond the specific problem of bipartite network projections, we believe that the perspective developed here has broader implications for statistical validation based on randomized ensembles and maximum-entropy null models. In all these approaches, different null models correspond to different statistical assumptions rather than simply different randomization procedures. Understanding how these assumptions affect the probability distribution of the test statistic is therefore essential for interpreting the outcome of hypothesis testing. From this viewpoint, the present work suggests that null models should be compared not only according to the structural constraints they preserve but also according to the statistical consequences that these constraints induce on the distribution of the quantities used for statistical inference.

The remainder of the paper is organized as follows. Section 2 introduces the four null models and discusses them as constrained statistical ensembles. Section 3 describes the three empirical bipartite systems and the statistical validation procedures. Section 4 analyses the first two moments of the co-occurrence distribution, derives the sparse approximation of the BiCM expectation, compares the statistically validated backbones obtained with the different null models and identifies when simplified analytical approaches accurately reproduce the microcanonical benchmark. Finally, Section 6 discusses the implications of our findings for the interpretation, comparison and future development of null-model-based statistical validation procedures.

\section{Null models and statistical validation framework}
\subsection{Statistical framework for null-model-based validation}

Consider a bipartite network composed of two disjoint sets of nodes,
denoted by $A$ and $B$, with $N_A$ and $N_B$ nodes, respectively.
The bipartite structure is represented by the binary matrix
$\mathbf{M}=\{m_{ij}\}$, where $m_{ij}=1$ if node $j$ of set $A$
is connected to node $i$ of set $B$, and $m_{ij}=0$ otherwise.
The degree of a node of type $A$ is denoted by $d_j$, whereas the
degree of a node of type $B$ is denoted by $k_i$.

The projection of the bipartite network onto set $A$ connects two
nodes whenever they share at least one common neighbour in set $B$.
The weight of the projected link is therefore naturally defined as
the number of common neighbours shared by the two nodes.
Throughout this work we denote by
\begin{equation}
X_{jj'}=\sum_{i=1}^{N_B}m_{ij}m_{ij'}
\label{eq:Xjj}
\end{equation}
the co-occurrence random variable associated with the pair of nodes
$j$ and $j'$ belonging to set $A$, and by
$X_{jj'}^{\mathrm{obs}}$ its value in the empirical bipartite network.

The central problem of statistical validation consists in assessing
whether the observed overlap $X_{jj'}^{\mathrm{obs}}$ is significantly
larger than expected under an appropriate null hypothesis.
This requires the specification of a null model capable of generating
the probability distribution

\begin{equation}
P(X_{jj'}),
\end{equation}

against which the empirical observation is compared.

Although the four null models investigated in this work differ in
their mathematical formulation, they all address the same statistical
problem: estimating the distribution of the random variable
$X_{jj'}$ under different assumptions on the structure of the
underlying bipartite network.
The probability that the observed overlap arises under the null
hypothesis is quantified through the one-sided $p$-value

\begin{equation}
\text{p-value}(X_{jj'}^{\mathrm{obs}})
=
P(X_{jj'}\ge X_{jj'}^{\mathrm{obs}}),
\label{eq:pvalueGeneral}
\end{equation}

which corresponds to the upper tail of the null distribution.

The statistical validation of the projected network is therefore
entirely determined by the probability distribution assigned to the
co-occurrence random variable.
Different null models produce different $p$-values because they
generate different probability distributions for $X_{jj'}$.
Consequently, differences among statistically validated backbones
should not be interpreted simply as differences between methodological approaches
but rather as the consequence of different statistical
assumptions encoded in the corresponding null hypotheses.

A complete analytical characterization of the distribution
$P(X_{jj'})$ is generally unavailable.
For the microcanonical ensemble the distribution must be estimated
numerically by repeated randomization, whereas for the analytical
models its exact evaluation may require approximations of different orders.
For this reason, an important part of the present work is devoted to
the analysis of the first two moments of the distribution,

\begin{equation}
\langle X_{jj'}\rangle,
\qquad
\mathrm{Var}(X_{jj'}),
\end{equation}

which represent the lowest-order statistical descriptors controlling
the position and the fluctuations of the null distribution.
As will be shown in the following sections, the combined behaviour of
these two quantities already explains most of the differences observed
among the statistically validated backbones obtained from the four
null models.

The four approaches considered in this work can be naturally organized
according to the structural information incorporated into the null
hypothesis.
The Curveball algorithm generates a microcanonical ensemble in which
the degree sequences of both node sets are preserved exactly.
The Bipartite Configuration Model (BiCM) relaxes these hard constraints
by preserving the same degree sequences only on average over a
canonical ensemble.
The Bipartite Partial Configuration Model (BiPCM) further simplifies
the description by preserving only the degree sequence of the
projected layer, while the Hypergeometric approach completely neglects
the degree heterogeneity of the non-projected layer.
This hierarchy of progressively relaxed constraints provides the
conceptual framework underlying the comparison developed throughout
the remainder of the paper.

\subsection{Microcanonical null model: the Curveball algorithm}

Among the null models considered in this work, the microcanonical
configuration model provides the most constrained statistical
description of the empirical bipartite network.
Its defining property is that the degree sequence of both node sets
is preserved exactly in every realization of the ensemble.
Consequently, all fluctuations observed within the ensemble originate
exclusively from the rewiring of links compatible with the empirical
degree sequences, while no fluctuations of node degrees are allowed.

The microcanonical ensemble therefore constitutes the natural
reference benchmark adopted throughout this work.
This choice should not be interpreted as identifying the
microcanonical ensemble with the ``ground truth'', since every null
model represents a statistical hypothesis rather than an objective
description of reality.
Instead, the microcanonical ensemble is used as a benchmark because,
among the considered null models, it preserves the largest amount of
empirical information by enforcing exactly the degree sequences of
both node sets.

The numerical generation of the microcanonical ensemble is performed
using the Curveball algorithm introduced by Strona \textit{et al.}
\cite{Strona2014}.
The algorithm generates unbiased random realizations of the
bipartite network while preserving exactly the row and column sums
of the empirical biadjacency matrix.
Compared with traditional swap-based randomization procedures,
Curveball achieves substantially faster mixing and has become one of
the standard methods for sampling microcanonical bipartite
configuration models.

For every realization of the ensemble, the co-occurrence between two
nodes $j$ and $j'$ belonging to the projected layer is computed as

\begin{equation}
X_{jj'}=\sum_{i=1}^{N_B}m_{ij}m_{ij'}.
\end{equation}

Repeating the randomization over a sufficiently large number of
independent realizations provides a numerical estimate of the
probability distribution

\begin{equation}
P_{\mathrm{MC}}(X_{jj'}),
\end{equation}

from which the cumulative distribution function is obtained.

Let $X_{jj'}^{\mathrm{obs}}$ denote the empirical co-occurrence
observed in the original bipartite network.
The statistical significance of the corresponding projected link is
evaluated through the upper-tail probability

\begin{equation}
\text{p-value}(X_{jj'}^{\mathrm{obs}})
=
P_{\mathrm{MC}}
\left(
X_{jj'}\ge
X_{jj'}^{\mathrm{obs}}
\right),
\label{eq:pvalueMC}
\end{equation}

which is estimated numerically from the sampled ensemble.

In practice, the cumulative distribution function is computed as

\begin{equation}
F(X_{jj'}^{\mathrm{obs}})
=
\frac{
\#\left(
X_{jj'}<
X_{jj'}^{\mathrm{obs}}
\right)
}
{N_{\mathrm{real}}},
\end{equation}

where $N_{\mathrm{real}}$ denotes the number of sampled
microcanonical realizations.
The corresponding one-sided $p$-value is therefore

\begin{equation}
\text{p-value}(X_{jj'}^{\mathrm{obs}})
=
1-
F(X_{jj'}^{\mathrm{obs}}).
\end{equation}

Since statistical validation involves a multiple-testing problem,
the numerical resolution of the estimated distribution must be
compatible with the adopted significance threshold after correcting
for multiple comparisons.
Throughout this work we employ the Bonferroni correction \cite{Hochberg1987}.
Consequently, the number of generated microcanonical realizations is
chosen to be at least equal to the ratio between the number of tested
projected links and the corrected significance level, ensuring that
the numerical estimation of the distribution possesses sufficient
resolution to discriminate statistically significant events.

The computational cost of the microcanonical approach increases
rapidly with network size because the probability distribution of
every projected link must be estimated numerically through repeated
randomization.
This computational burden constitutes one main motivation for
the development of analytical approximations such as the BiCM,
BiPCM and Hypergeometric models discussed in the following
subsections.

\subsection{Canonical null models}

The microcanonical ensemble discussed in the previous subsection
preserves exactly the degree sequence of both node sets in every
realization of the ensemble.
Although this provides the most constrained statistical description
of the empirical bipartite network, the numerical estimation of the
co-occurrence distribution becomes computationally demanding for
large systems.
A natural alternative is provided by canonical ensembles, in which
the degree constraints are imposed only on ensemble averages.
Within the framework of maximum-entropy statistical mechanics, this
corresponds to replacing hard constraints by soft constraints,
thereby allowing fluctuations of the degree sequence from one
realization to another while preserving its expected value.

The random bipartite networks belonging to the canonical ensemble
are generated from a probability matrix
$\mathbf{P}=\{p_{ij}\}$,
whose element $p_{ij}$ denotes the probability that node $i$ of
set $B$ is connected to node $j$ of set $A$.
Once the probability matrix has been determined,
a realization of the ensemble is obtained by independently drawing
each entry of the biadjacency matrix according to

\begin{equation}
b_{ij}=
\left\{
\begin{array}{ll}
1, & u_{ij}\le p_{ij},\\
0, & u_{ij}>p_{ij},
\end{array}
\right.
\label{eq:fillingScheme}
\end{equation}

where $u_{ij}$ is a random number uniformly distributed in the
interval $(0,1)$.

The different canonical null models considered in this work differ
only in the constraints imposed on the probability matrix
$\mathbf{P}$.
Consequently, they generate different probability distributions for
the co-occurrence random variable $X_{jj'}$.

\subsubsection{Bipartite Configuration Model}

The Bipartite Configuration Model (BiCM)
\cite{Saracco2015,Saracco2017}
constitutes the canonical counterpart of the microcanonical
configuration model.
Its defining feature is that the expected degree of every node in
both layers is constrained to reproduce the empirical degree
sequence,

\begin{equation}
\sum_{i=1}^{N_B}p_{ij}=d_j,
\qquad
\forall j,
\end{equation}

and

\begin{equation}
\sum_{j=1}^{N_A}p_{ij}=k_i,
\qquad
\forall i.
\end{equation}

The probability matrix is obtained through the
maximum-likelihood solution of the corresponding
maximum-entropy ensemble, as described in
Refs.~\cite{Saracco2015,Saracco2017}.
Unlike the microcanonical ensemble, individual realizations do not
preserve the empirical degree sequence exactly.
Instead, node degrees fluctuate around their empirical values while
remaining unbiased over the ensemble.

For a given pair of nodes $j$ and $j'$ belonging to the projected
layer, the probability that a node $i$ of the opposite layer is
connected simultaneously to both nodes is

\begin{equation}
P(V_{jj'}^{i}=1)=p_{ij}p_{ij'},
\end{equation}

where $V_{jj'}^{i}$ denotes the corresponding V-motif.
The co-occurrence random variable is therefore

\begin{equation}
X_{jj'}
=
\sum_{i=1}^{N_B}
V_{jj'}^{i},
\end{equation}

namely the sum of independent but non-identically distributed
Bernoulli random variables.
Consequently, $X_{jj'}$ follows a Poisson--Binomial distribution.

The exact probability mass function can be evaluated numerically
through the Fourier-transform algorithm proposed by
Hong~\cite{Hong2013}, allowing an analytical computation of
the $p$-values without explicitly generating a canonical ensemble of
random bipartite networks.
Throughout the present work we employ the exact
Poisson--Binomial distribution rather than the asymptotic
approximations proposed in
Ref.~\cite{Saracco2017}, since the former provides the exact
analytical distribution of the co-occurrence.

\subsubsection{Bipartite Partial Configuration Model}

The Bipartite Partial Configuration Model (BiPCM)
\cite{Saracco2017,Neal2024}
represents a further relaxation of the canonical description.
Instead of preserving the expected degree sequence of both node
sets, only the expected degree sequence of the projected layer is
constrained, while the opposite layer is replaced by a homogeneous
statistical description.

Under this assumption, the connection probability becomes

\begin{equation}
p_{ij}
=
\frac{d_j}{N_B},
\label{eq:BiPCMprob}
\end{equation}

which is independent of the degree of node $i$ belonging to the
non-projected layer.
Consequently, all nodes of type $B$ become statistically equivalent.

This simplification has two important consequences.
First, the expected degree of every node of type $B$ is identical,

\begin{equation}
\langle k_i\rangle
=
\frac{L}{N_B},
\end{equation}

where $L$ denotes the total number of links in the bipartite
network.
Second, the variance of the degree is also identical for all nodes,

\begin{equation}
\mathrm{Var}(k_i)
=
\sum_{j=1}^{N_A}
\frac{d_j}{N_B}
\left(
1-
\frac{d_j}{N_B}
\right),
\end{equation}

showing that the heterogeneity of the non-projected layer is no
longer preserved.
The degree distribution of nodes of type $B$ is therefore entirely
determined by the degree sequence of the projected layer.

The BiPCM thus occupies an intermediate position between the BiCM
and the Hypergeometric approximation discussed in the next
subsection.
While it retains a canonical description of the projected layer, it
replaces the detailed heterogeneity of the opposite layer by its
average statistical properties.
As will be shown in Section~4, this approximation has important
consequences for both the expectation and the variance of the
co-occurrence distribution.

\subsection{Hypergeometric approximation}

The Hypergeometric approach represents the simplest statistical
description considered in this work \cite{Tumminello2011}.
Unlike the canonical null models discussed in the previous subsection,
it does not attempt to preserve the heterogeneity of the
non-projected layer.
Instead, it assumes that the neighbours of every node in the projected
layer are selected uniformly at random among the $N_B$ nodes of the
opposite layer.
Consequently, only the degree sequence of the projected layer enters
the statistical description.

Within this framework, the probability that two nodes
$j$ and $j'$ of the projected layer, having degrees
$d_j$ and $d_{j'}$, share exactly $X$ common neighbours
is given by the Hypergeometric distribution

\begin{equation}
H(X|N_B,d_j,d_{j'})
=
\frac{
\binom{d_j}{X}
\binom{N_B-d_j}{d_{j'}-X}
}
{
\binom{N_B}{d_{j'}}
},
\label{eq:Hyper}
\end{equation}

which follows directly from the classical urn problem.
The statistical significance of the observed overlap
$X_{jj'}^{\mathrm{obs}}$ is therefore obtained from the
upper tail of the distribution,

\begin{equation}
\text{p-value}(X_{jj'}^{\mathrm{obs}})
=
\sum_{X\ge X_{jj'}^{\mathrm{obs}}}
H(X|N_B,d_j,d_{j'}).
\label{eq:HyperPvalue}
\end{equation}

An important property of the Hypergeometric model is that it
preserves exactly the observed degree of every node belonging to the
projected layer while completely neglecting the heterogeneity of the
non-projected layer.
For this reason, the Hypergeometric distribution may be interpreted
as the microcanonical counterpart of the Bipartite Partial
Configuration Model.
Indeed, both models share the same expected co-occurrence,

\begin{equation}
\langle X_{jj'}\rangle
=
\frac{d_jd_{j'}}{N_B},
\label{eq:HyperMean}
\end{equation}

although they differ in the fluctuations around this expectation,
since the Hypergeometric model imposes hard constraints on the degree
sequence of the projected layer whereas the BiPCM preserves these
constraints only on average.

From the perspective adopted in the present work, the Hypergeometric
model constitutes the limiting case of the hierarchy of null models.
It represents the strongest analytical simplification because all
information concerning the degree heterogeneity of the
non-projected layer is discarded.
Nevertheless, as will be shown in Section~4, despite this drastic
simplification the Hypergeometric approximation reproduces some
statistical properties of the microcanonical ensemble remarkably
well. Understanding the origin of this behaviour constitutes one of
the central questions addressed in this work.

\subsection{Hierarchy of null models and statistical assumptions}

The four null models considered in this work should not be regarded as
independent statistical procedures. Rather, they define a hierarchy of
statistical descriptions obtained by progressively relaxing the
constraints imposed on the empirical bipartite network. This hierarchy
provides the conceptual framework underlying the comparison developed
throughout the remainder of the paper.

At the highest level of constraint lies the microcanonical
configuration model, sampled through the Curveball algorithm.
In this ensemble the degree sequence of both node sets is preserved
exactly in every realization. Consequently, all observed fluctuations
originate exclusively from the rewiring of links compatible with the
empirical degree sequences.

The Bipartite Configuration Model (BiCM) constitutes the canonical
counterpart of the microcanonical ensemble. Here the same degree
sequences are no longer imposed exactly but only in expectation over
the ensemble. The resulting fluctuations of node degrees generate a
different probability distribution for the co-occurrence random
variable while substantially reducing the computational cost of
the statistical validation.

The Bipartite Partial Configuration Model (BiPCM) represents a further
relaxation of the statistical constraints. While preserving the
expected degree sequence of the projected layer, it replaces the
opposite layer by a homogeneous statistical description.
Consequently, the detailed heterogeneity of the non-projected layer is
no longer incorporated into the null hypothesis.

Finally, the Hypergeometric model represents the limiting case of this
hierarchy. It preserves exactly the degree sequence of the projected
layer but completely neglects the heterogeneity of the opposite layer.
Its simplicity allows an analytical computation of the probability
distribution of the co-occurrence while requiring only the degree
sequence of the projected nodes.

The progressive relaxation of constraints is accompanied by an
increasing degree of analytical tractability. The microcanonical model
requires repeated randomization of the bipartite network in order to
estimate numerically the probability distribution of the co-occurrence.
The BiCM and BiPCM replace the numerical randomization by analytical
descriptions based on Poisson--Binomial or Binomial statistics, whereas the
Hypergeometric model provides a closed-form analytical expression for
the probability distribution.

This hierarchy therefore represents
a trade-off between statistical fidelity and computational efficiency.
More constrained ensembles preserve a larger amount of empirical
information but require increasingly demanding numerical procedures.
Conversely, simpler analytical models reduce the computational burden
at the cost of progressively stronger approximations of the underlying
bipartite structure.

The central hypothesis investigated in the present work is that these
different statistical assumptions manifest themselves through the
probability distribution of the co-occurrence random variable
$X_{jj'}$. Since the complete distribution is generally unavailable in
closed analytical form, we focus on its first two moments, namely the
expected co-occurrence and its variance. As will be shown in the
following sections, the ability of a null model to reproduce these two
quantities largely determines its capability to recover the
statistically validated backbone obtained from the micro canonical
benchmark.

Table~\ref{tab:nullmodels} summarizes the principal characteristics of
the four null models considered in this work.

\begin{table}[t]
\centering
\caption{Conceptual comparison of the four null models investigated in
this work.}
\label{tab:nullmodels}
\begin{tabular}{lcccc}
\hline
Model &
Ensemble &
Constraints &
Distribution &
Computational\\

~ &
~ &
~ &
~ &
cost\\

\hline
Curveball &
Microcanonical &
Exact degrees of $A$ and $B$ &
Numerical &
Very High\\

BiCM &
Canonical &
Expected degrees of $A$ and $B$ &
Poisson--Binomial &
Moderate\\

BiPCM &
Canonical &
Expected degrees of $A$ only &
Binomial &
Low\\

Hypergeometric &
Urn model &
Exact degrees of $A$ only &
Hypergeometric &
Low\\
\hline
\end{tabular}
\end{table}

\section{Empirical bipartite systems}

The methodology developed in this work is assessed on three empirical
bipartite systems originating from different scientific domains,
namely comparative genomics, international trade and food science.
Besides representing unrelated application fields, these datasets span
substantially different regimes of network density and degree
heterogeneity, making them particularly suitable for investigating the
effect of different null-model assumptions on the statistical
validation of projected bipartite networks.

In all three cases the original system is naturally represented as a
bipartite network composed of two disjoint node sets, denoted by $A$
and $B$. Statistical validation is always performed on the projection
onto set $A$, while the nodes of set $B$ mediate the observed
co-occurrences. The three projected networks correspond respectively
to organisms, countries and ingredients.

\subsection{COG database: organisms projected network}

The first dataset is obtained from the Clusters of Orthologous Genes
(COG) database
\cite{Tatusov1997,Tatusov2003}.
The database describes the occurrence of orthologous gene families in
fully sequenced genomes and naturally defines a bipartite network
whose two node sets are organisms (set $A$) and COGs (set $B$).

The 2003 release considered in this work contains
$N_A=66$ organisms (13 Archaea, 50 Bacteria and 3 unicellular
Eukaryota) and $N_B=4873$ distinct COGs.
Additional metadata are available for both node sets, including the
taxonomic classification of organisms and the functional category of
each COG.

The projection onto the organism layer connects two organisms whenever
they share at least one COG.
The weight of each projected link is given by the number of common
COGs.
The resulting projected network is complete, containing
66 nodes and all
$66(66-1)/2=2145$ possible links.

\subsection{World Trade Web: countries projected network}

The second dataset is derived from the World Trade Web (WTW),
using international trade data extracted from the BACI
database
\cite{CEPII}. Specifically, we consider the data referring to the 2023 calendar year.
Following the preprocessing procedure proposed by
Saracco \textit{et al.}~\cite{Saracco2017},
products originally classified at the six-digit level are aggregated
to four-digit product classes.
Country--product associations are retained only when the corresponding
Revealed Comparative Advantage (RCA) satisfies
$\mathrm{RCA}\ge1$.

The resulting bipartite network consists of
$N_A=226$ countries and
$N_B=1348$ products.

The projection onto the country layer connects two countries whenever
they export at least one common product.
The weight of the projected link corresponds to the number of shared
products.
The resulting projected network contains one connected component with
226 nodes and 24\,238 links, corresponding to approximately
95\% of all possible links.

\subsection{CulinaryDB: ingredients projected network}

The third dataset is obtained from CulinaryDB
\cite{Culinary}
which records the ingredients composing culinary recipes.

We restrict the dataset to recipes satisfying the so-called
``rule of five'' \cite{Oliver2017}, namely recipes composed of five principal
ingredients and, in addition, up to four ubiquitous ingredients such as water,
salt, pepper and vinegar.
The resulting bipartite network consists of
$N_A=508$ ingredients and
$N_B=4454$ recipes.

The projected network is obtained by connecting two ingredients
whenever they appear together in at least one recipe.
The weight of each projected edge equals the number of recipes
containing both ingredients.
Unlike the previous two datasets, the resulting projected network is
relatively sparse, containing only 12\,455 links out of the
128\,778 possible connections.

\subsection{Structural characteristics of the datasets}

The three empirical systems provide complementary structural regimes
for assessing the performance of the different null models.
The COG network represents a dense projection generated by a highly
heterogeneous non-projected layer.
The World Trade Web occupies an intermediate regime, whereas the
CulinaryDB projection is substantially sparser and corresponds to a
bipartite system that closely satisfies the assumptions underlying the
sparse analytical approximations discussed in Section~4.

As will be shown later, these structural differences play a central
role in determining the agreement between the different null models.
Rather than simply providing examples from distinct application
domains, the three datasets allow us to investigate how network
density and degree heterogeneity influence the statistical validation
of projected bipartite networks.

\section{Results}

We use the statistically validated links obtained from the microcanonical bipartite configuration model, sampled through the Curveball algorithm, as the reference benchmark for assessing the performance of three analytical null models: the Bipartite Configuration Model (BiCM), the Bipartite Partial Configuration Model (BiPCM), and the Hypergeometric model (H). The comparison is performed on three bipartite systems: the COG network, with organisms in set $A$ and clusters of orthologous genes in set $B$; the World Trade Web (WTW) for 2023, with countries in set $A$ and traded products in set $B$; and CulinaryDB, with ingredients in set $A$ and recipes in set $B$. These datasets span markedly different regimes of degree heterogeneity in the non-projected layer $B$, from the strongly heterogeneous COG network to the weakly heterogeneous CulinaryDB network, with WTW displaying an intermediate behavior.

All links are validated at significance level $\alpha=0.05$ after Bonferroni correction. Table~\ref{table X} summarizes the main structural properties of the three bipartite networks and reports the number of statistically validated links obtained under the microcanonical benchmark.

\begin{table}[H]
\centering
\caption{Summary of main indicators of the bipartite networks and of the set A projected network. Density gives the percent of link density. $\rho=\frac{\sigma_B^2}{\langle k_B\rangle^2}$ is the square of the coefficient of variation of degree of nodes of set B. The last column gives the number of statistically validated links of projected network of set A obtained according to the microcanonical (M) null models with statistical threshold $\alpha=0.05$ and Bonferroni correction.}
\label{table X}
\scalebox{0.8}{
\begin{tabular}{c | c c c c c c c c c | c}
\toprule
\textbf{Dataset} & Node~type & Node~type & $Nodes$ & $Nodes$ & $Bipartite$ & $Bipartite$ & $Set~A$ & $Set~A$ & $Set~B$ & $M$ \\
~ & Set~A & Set~B & $set~A$ & $set~B$ & $links$ & $density$ & $proj~links$ & $density$ & $\rho$ & $val~links$ \\
\midrule
COG 2003  & Organisms & COGs & 66 & 4873 & 83675 & 26.0 \% & 2145 & 100 \% & 0.867  & 882 \\
WTW 2023 & Countries & Products & 226 & 1348 & 32815 & 10.8 \% & 24238 & 95.3 \% & 0.264 & 672 \\
Culinary DB & Ingredients & Recipes & 508 & 4454 & 26129 & 1.15 \% & 12455 & 9.67 \% & 0.0207 & 373 \\
\bottomrule
\end{tabular}}
\end{table}

\subsection{Recovery of the microcanonical backbone}

The three analytical null models exhibit distinct patterns of agreement with the microcanonical benchmark (Table~\ref{tab:Y}). BiCM is systematically conservative: in all three datasets it produces no false positives, and therefore reaches unit precision and specificity, but it misses a substantial fraction of the links validated by Curveball. Its recall is $0.685$ for COG, $0.433$ for WTW, and $0.611$ for CulinaryDB.

BiPCM and H behave differently. For the heterogeneous COG and WTW networks, both methods recover all Curveball-validated links, yielding unit recall and negative predictive value, but they also validate a large number of additional links. This produces low precision and, especially for COG, low specificity. In CulinaryDB, by contrast, neither BiPCM nor H introduces false positives. Their main limitation is instead the presence of false negatives, with H achieving the highest recall among the three analytical models.

These results indicate that the performance of a null model cannot be characterized by a single scalar measure. Rather, it reflects the combined ability of the model to locate and shape the null distribution of the pairwise co-occurrence $X_{ij}$. We therefore next compare the first two moments of this distribution.

\begin{table}[H]
\centering
\caption{Number of statistically validated links according to different null models: microcanonical (M), canonical (BiCM), partial canonical (BiPCM), and Hypergeometric (H). By using minrocanonical validated links as a benchmark, we also indicates the number of predicted positive (PP) and predicted negative (PN) links for the models BiCM, BiPCM and H. For each model the PP are split into true positive (TP) and false positive (FP) and the PN are split into true negative (TN) and false negative (FN). The bottom rows of the table reports the Recall (~TP/(TP+FN)~), Precision (~TP/(TP+FP)~), Specificity (~TN/(TN+FP)~) and negative predictive value (~TN/(TN+FN)~).}
\label{tab:Y}
\scalebox{0.66}{
\begin{tabular}{c | c | c c c c | c c c c | c c c c}
\toprule
Dataset & M & \multicolumn{4}{c|}{BiCM} & \multicolumn{4}{c|}{BiPCM} & \multicolumn{4}{c}{H} \\
\hline
~ & ~ & $PP$ & $TP$ & $PN$ & $TN$ & $PP$ & $TP$ & $PN$ & $TN$ & $PP$ & $TP$ & $PN$ & $TN$ \\
~ & ~ & ~ & $FP$ & ~ & $FN$ & ~ & $FP$ & ~ & $FN$ & ~ & $FP$ & ~ & $FN$ \\
\midrule
COG 2003  & 882 & 604 & 604 & 1541 & 1263 & 2021 & 882 & 124 & 124 & 2067 & 882 & 78 & 78 \\
~  & ~ & ~ & 0 & ~ & 278 & ~ & 1139 & ~ & 0 & ~ & 1185 & ~ & 0 \\
\hline
WTW 2023 & 672 & 291 & 291 & 23566 & 23185 & 907 & 672 & 23331 & 23331 & 1532 & 672 & 22706 & 22706 \\
~  & ~ & ~ & 0 & ~ & 381 & ~ & 235 & ~ & 0 & ~ & 860 & ~ & 0 \\
\hline
Culinary & 373 & 228 & 228 & 12082 & 11937 & 236 & 236 & 12219 & 12082 & 293 & 293 & 12162 & 12082 \\
~  & ~ & ~ & 0 & ~ & 145 & ~ & 0 & ~ & 137 & ~ & 0 & ~ & 80 \\
\hline
Dataset & ~ & \multicolumn{4}{c|}{BiCM} & \multicolumn{4}{c|}{BiPCM} & \multicolumn{4}{c}{H} \\
\hline
~ & ~ & $Rec$ & $Pre$ & $Spe$ & $NPV$ & $Rec$ & $Pre$ & $Spe$ & $NPV$ & $Rec$ & $Pre$ & $Spe$ & $NPV$ \\
\hline
COG 2003  & ~ & 0.685 & 1.00 & 1.00 & 0.820 & 1.00 & 0.436 & 0.0982 & 1.00 & 1.00 & 0.427 & 0.0618 & 1.00 \\
\hline
WTW 2023 & ~ & 0.433 & 1.00 & 1.00 & 0.984 & 1.00 & 0.741 & 0.990 & 1.00 & 1.00 & 0.439 & 0.963 & 1.00 \\
\hline
Culinary & ~ & 0.611 & 1.00 & 1.00 & 0.988 & 0.633 & 1.00 & 1.00 & 0.989 & 0.785 & 1.00 & 1.00 & 0.994 \\
\bottomrule
\end{tabular}}
\end{table}

\subsection{Mean and variance of the co-occurrence distribution}

\begin{figure}[H]
\centering
\includegraphics[width=\textwidth]{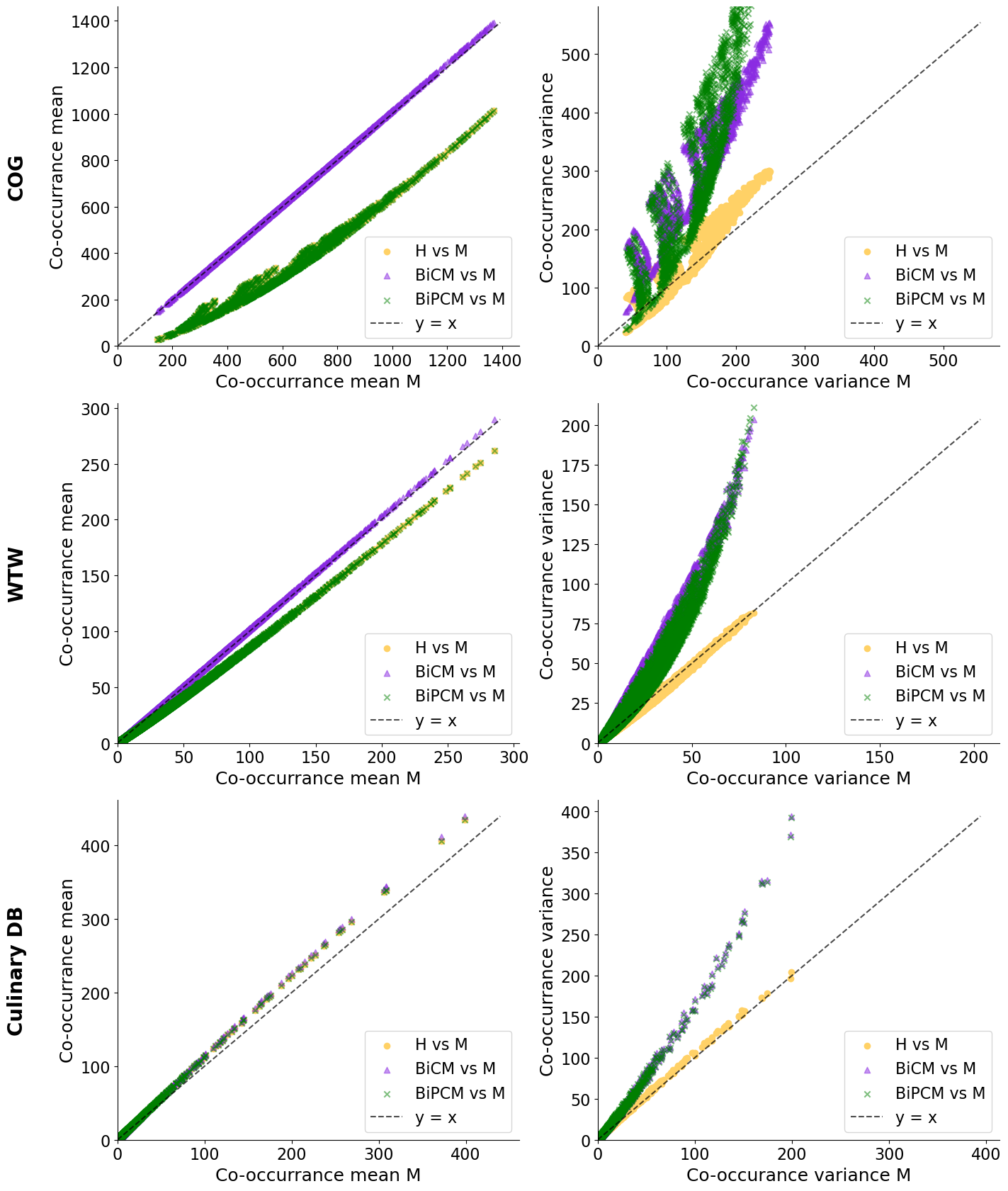}
\caption{Scatter plots comparing the expected value $\langle X_{jj'}\rangle$ (left panels) and the variance $\mathrm{Var}\{X_{jj'}\}$ (right panels) obtained for the BiCM (purple triangles) , BiPCM (green crosses), and Hypergeometric (orange circles) null models with the values numerically estimated for the microcanonical null model for the COG (top row), WTW (middle row), and culinary (bottom row) datasets. The dashed line is $y=x$.}
\label{fig:A}
\end{figure}

Figure~\ref{fig:A} compares, for every pair of nodes in the projected layer, the expected value $\langle X_{jj'}\rangle$ and variance $\mathrm{Var}(X_{jj'})$ predicted by BiCM, BiPCM, and H with the corresponding quantities estimated numerically under the microcanonical model.

For COG, BiCM reproduces the microcanonical expectation substantially better than BiPCM and H. The latter two models systematically underestimate $\langle X_{jj'}\rangle$, consistent with their neglect of the heterogeneity of layer $B$. At the same time, BiCM and BiPCM predict variances larger than those observed under Curveball, whereas H provides the closest approximation to the microcanonical variance. These two errors act in opposite directions. Underestimating the mean makes the empirical co-occurrence appear more exceptional and therefore increases the number of positive validations. Inflating the variance broadens the null distribution and reduces statistical significance. This explains why BiPCM and H attain high recall but low precision in COG, whereas BiCM attains perfect precision at the cost of many false negatives.

WTW exhibits the same qualitative mechanism. BiCM again provides the most accurate estimate of the mean, while BiPCM and H underestimate it. Conversely, H reproduces the microcanonical variance more closely than the two canonical models, whose variances are systematically larger. The overall validation performance therefore reflects a balance between a mean bias that promotes positive predictions and a variance inflation that suppresses them.

CulinaryDB displays a different regime. All three analytical models slightly overestimate the microcanonical expectation. This shift reduces the distance between the observed overlap and the center of the null distribution and therefore acts in the direction of decreasing statistical significance. BiCM and BiPCM additionally overestimate the variance, so that both the mean and variance errors favor false negatives. H performs best because, although its mean is also slightly biased upward, its variance remains close to the microcanonical value.

These observations show that neither the mean nor the variance alone determines the recovered backbone. Statistical validation is controlled by their joint effect on the upper tail of the null distribution.

\subsection{Sparse-limit approximation for the BiCM expectation}

For BiPCM and H, the expected co-occurrence is
\begin{equation}
\langle X_{jj'}\rangle_{\mathrm{BiPCM}}
=
\langle X_{jj'}\rangle_{\mathrm{H}}
=
\frac{d_j d_{j'}}{N_B},
\label{eq:mean_h_bipcm}
\end{equation}
where $d_j$ and $d_{j'}$ are the degrees of nodes $j$ and $j'$ in the projected layer $A$, and $N_B$ is the number of nodes in the non-projected layer.

For BiCM, the exact expectation is
\begin{equation}
\langle X_{jj'}\rangle_{\mathrm{BiCM}}
=
\sum_{\ell=1}^{N_B} p_{\ell j}p_{\ell j'},
\label{eq:mean_bicm_exact}
\end{equation}
where $p_{\ell j}$ is the BiCM probability of observing the bipartite edge $(\ell,j)$. In the sparse regime, the BiCM connection probability can be linearized as
\begin{equation}
p_{\ell j}\simeq \frac{k_{\ell}d_j}{L},
\label{eq:sparse_p}
\end{equation}
where $k_{\ell}$ is the degree of node $\ell\in B$ and $L$ is the total number of bipartite links. Substituting Eq.~\eqref{eq:sparse_p} into Eq.~\eqref{eq:mean_bicm_exact} gives
\begin{equation}
\langle X_{jj'}\rangle_{\mathrm{BiCM}} \simeq
\frac{d_j d_{j'}}{L^2}
\sum_{\ell=1}^{N_B}k_{\ell}^2 =
\frac{d_j d_{j'}}{N_B}
\frac{\langle k_B^2\rangle}{\langle k_B\rangle^2} =
\frac{d_j d_{j'}}{N_B}
\left(1+\frac{\sigma_B^2}{\langle k_B\rangle^2}\right).
\label{eq:mean_bicm_sparse}
\end{equation}

Here $\langle k_B\rangle$ and $\sigma_B^2$ denote the mean and variance of the degree distribution of layer $B$. Equation~\eqref{eq:mean_bicm_sparse} decomposes the BiCM expectation into the homogeneous contribution shared by BiPCM and H and a correction proportional to the squared coefficient of variation,
\begin{equation}
\rho=\frac{\sigma_B^2}{\langle k_B\rangle^2}.
\end{equation}
Thus, within the sparse approximation, neglecting the degree heterogeneity of the non-projected layer induces a downward bias in the expected co-occurrence whose relative magnitude is controlled by $\rho$.

\subsection{Domain of validity of the sparse approximation}

\begin{figure}[H]
\centering
\includegraphics[width=\textwidth]{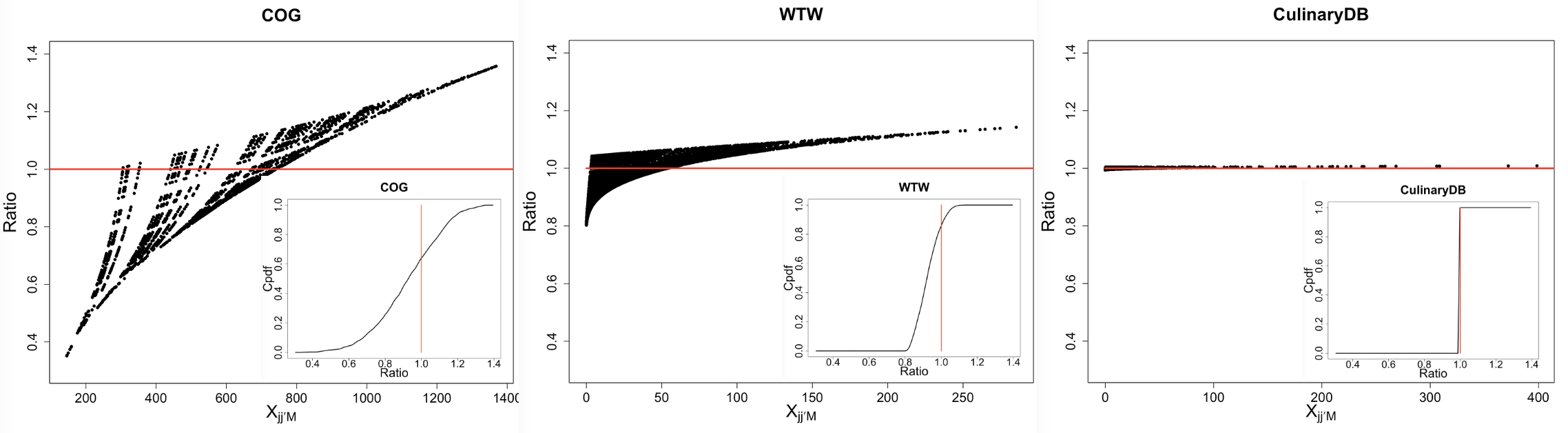}
\caption{Ratio between the approximated value of the expected co-occurrence estimated by Eq. \ref{eq:mean_bicm_sparse} divide by the BiCM expectation obtained by using the Poisson-Binomial distribution against the microcanonical expectation of co-occurrence for all pairs of Set A nodes. The three panels refer to COG (left), WTW (middle), and CulinaryDB (right).The insets show the cumulative probability density function of the ratio values. They are located around the value $R=1$ highlighted by the red line.}
\label{fig:sparse}
\end{figure}

In Fig.~\ref{fig:sparse} we show the ratio between the approximated value of the expected co-occurrence estimated by Eq. \ref{eq:mean_bicm_sparse} divided by the BiCM co-occurrence expectation obtained by using the Poisson-Binomial distribution . The approximation is not equally accurate for the three datasets. In the COG dataset, a substantial fraction of pairs significantly deviates from the linearized probability approximation (i.e., from $Ratio=1$ indicated as a red line in the figure). In WTW, violations are less pronounced but still sizeable  whereas in  CulinaryDB, we verify an excellent accuracy of the sparse-limit formula. In all cases ratio values crosses the value $Ratio=1$ (see insets in the figure showing the cumulative probability density function of the $Ratio$).

This ordering is consistent with the condition underlying Eq.~\eqref{eq:sparse_p}. The approximation requires $k_{\ell}d_j/L<1$ for all pairs of nodes. As a direct diagnostic, we computed the fraction of bipartite node pairs for which $k_{\ell}d_i/L>1$. This fraction is $0.12$ for COG, $0.0057$ for WTW, and zero for CulinaryDB. 

The sparse expression therefore works for the limit of very sparse networks and should be interpreted as a regime-dependent analytical approximation. It correctly identifies the structural contribution of degree heterogeneity in layer $B$, but its quantitative accuracy depends on the validity of the sparse linearization.

\subsection{Analytical comparison of the variances}

BiPCM and H share the same expectation but differ in their variance. Their analytical expressions are
\begin{equation}
\mathrm{Var}_{\mathrm{BiPCM}}(X_{jj'})
=
\frac{d_j d_{j'}}{N_B}
\left(1-\frac{d_j d_{j'}}{N_B^2}\right),
\end{equation}
and
\begin{equation}
\mathrm{Var}_{\mathrm{H}}(X_{jj'})
=
\frac{d_j d_{j'}}{N_B}
\frac{(N_B-d_j)(N_B-d_{j'})}{N_B(N_B-1)}.
\end{equation}
For $N_B>2$ and $0\le d_i,d_j\le N_B$, one obtains
\begin{equation}
\mathrm{Var}_{\mathrm{H}}(X_{jj'})
\le
\mathrm{Var}_{\mathrm{BiPCM}}(X_{jj'}).
\end{equation}
Hence, at fixed mean, BiPCM is intrinsically more dispersed than H. This analytical result is consistent with the numerical comparison in Fig.~\ref{fig:A}. For BiCM, the variance is computed numerically; in all three datasets it is also larger than the Hypergeometric variance for the considered pairs.

\begin{table}[H]
\centering
\caption{Number of statistically validated links according to the ``Hybrid CH" null model. We use microcanonical validated links as a benchmark. We indicate the number of predicted positive (PP) and predicted negative (PN) links for the Hybrid CH null model. The bottom rows of the table reports the Recall (~TP/(TP+FN)~), Precision (~TP/(TP+FP)~), Specificity (~TN/(TN+FP)~) and negative predictive value (~TN/(TN+FN)~). We also report the square of the coefficient of variation $CV=\sigma_B/\langle d_B\rangle$, the number of links of the microcanonical benchmark (M) and the F1 indicator (F1=2 x Recall x Precision/(Recall + Precision)).}
\label{table K}
\scalebox{0.85}{
\begin{tabular}{c | c | c | c | c c c c}
\toprule
Dataset & $CV^2$ & Set A & M & \multicolumn{4}{c}{Hybrid CH} \\
~ & ~ & projected links & validate links & ~ & ~ & ~ & ~ \\
\hline
~ & ~ & ~ & ~ & $PP$ & $TP$ & $PN$ & $TN$ \\
~ & ~ & ~ & ~ & ~ & $FP$ & ~ & $FN$ \\
\hline\hline
COG 2003 & 0.8671 & 2145 & 882 & 761 & 761 & 1384 & 1263 \\
~ & ~ & ~ & ~ & ~ & 0 & ~ & 121 \\
\hline
WTW 2023 & 0.2643 & 24238 & 672 & 751 & 651 & 23487 & 23466 \\
~ & ~ & ~ & ~ & ~ & 100 & ~ & 21 \\
\hline
Culinary & 0.0207 & 12455 & 373 & 291 & 291 & 12455 & 12373 \\
~ & ~ & ~ & ~ & ~ & 0 & ~ & 82 \\
\hline\hline
Dataset & ~ & ~ & \multicolumn{5}{c}{Hybrid CH} \\
\hline
~ & ~ & ~ & $F1$ & $Rec$ & $Pre$ & $Spe$ & $NPV$ \\
\hline
COG 2003 & ~ & ~ & 0.926 & 0.863 & 1.00 & 1.00 & 0.912 \\
\hline
WTW 2023 & ~ & ~ & 0.915 & 0.969 & 0.867 & 0.996 & 0.999 \\
\hline
Culinary & ~ & ~ & 0.876 & 0.780 & 1.00 & 1.00 & 0.993 \\
\bottomrule
\end{tabular}}
\end{table}

\subsection{A hybrid approximation}

The previous analysis suggests that the most accurate components of the analytical descriptions are model-dependent. BiCM generally provides the best approximation to the microcanonical co-occurrence mean in COG and WTW, whereas H provides the closest approximation to the microcanonical co-occurrence variance across the three datasets. This observation motivates a hybrid approximation that combines the BiCM expectation with the Hypergeometric variance.

Table~\ref{table K} reports the links validated by this Hybrid CH and its performance relative to the microcanonical benchmark. The hybrid model improves the balance between recall and precision for COG and WTW, yielding $F_1$ scores of $0.926$ and $0.915$, respectively. For CulinaryDB, its performance is comparable to that of H, consistent with the fact that the mean expectations of the analytical models are already close in this weakly heterogeneous regime.

These results indicate that the discrepancies among validation methods can be traced to identifiable errors in the location and width of the null distribution. Correcting these components separately provides a computationally tractable approximation with high fidelity to the microcanonical benchmark. At the same time, the sparse-limit analysis clarifies when the heterogeneity correction to the co-occurrence mean can be used quantitatively and when the full BiCM expectation remains necessary.

\section{Discussion}

The statistical validation of projected bipartite networks relies fundamentally on the specification of an appropriate null model. Although different null models have been proposed in the literature, their comparison has often focused on the resulting statistically validated backbones rather than on the statistical assumptions that generate them. The present work adopts a different perspective. Instead of asking which null model performs best, we investigate how the structural constraints defining the null model influence the probability distribution of the co-occurrence random variable and how these modifications propagate into the outcome of statistical validation.

Our results show that the choice between microcanonical, canonical and simplified analytical null models cannot be interpreted simply as a choice between different methodological approaches. Each null model corresponds to a different statistical description of the underlying bipartite system, preserving different amounts of structural information and therefore generating different probability distributions for the co-occurrence $X_{jj'}$. Consequently, different statistically validated backbones should not be regarded as contradictory results but as the natural consequence of different statistical hypotheses.

A central outcome of our analysis is that the comparison of null models should be performed not only in terms of the explicit constraints they preserve but also in terms of the statistical consequences that these constraints induce on the null distribution of the test statistic. While the complete probability distribution of the co-occurrence is analytically inaccessible for most models, our results indicate that much of the observed behaviour can already be understood through the first two moments of the distribution. The expectation determines the position of the null distribution, whereas the variance determines the magnitude of its fluctuations. Together, these two quantities largely explain the observed differences in true positives, false positives and false negatives among the four validation procedures investigated.

The three empirical datasets considered in this work illustrate that the effects of relaxing the null-model constraints are not universal but depend on the structural properties of the underlying bipartite network. In the COG and WTW datasets, neglecting the heterogeneity of the non-projected layer produces a substantial bias in the expected overlap, thereby increasing the number of statistically validated links and generating additional false positives. Conversely, the larger variances predicted by the canonical models broaden the null distribution, reducing statistical significance and producing a more conservative validation characterized by an increased number of false negatives. In the CulinaryDB network the situation is different. Here all analytical models slightly overestimate the expected overlap, while the Hypergeometric approximation reproduces remarkably well the variance of the microcanonical ensemble. As a consequence, the mean and variance contribute differently to the final validation outcome than in the other two systems. These observations demonstrate that the statistical effects of relaxing the constraints cannot be inferred a priori but must be evaluated for the specific class of networks under investigation.

An important theoretical result of the present work is the derivation of a leading-order sparse approximation for the expected overlap under the Bipartite Configuration Model. Although this approximation becomes quantitatively accurate only in sufficiently sparse bipartite networks, it identifies the normalized degree heterogeneity of the non-projected layer,
$\rho=\frac{\sigma_B^2}{\langle k_B\rangle^2}$ , as the leading correction to the Hypergeometric expectation. In this sense, the approximation should not primarily be regarded as a universally accurate estimator of the expected overlap, but rather as an analytical tool that isolates the dominant structural mechanism through which degree heterogeneity affects the first moment of the null distribution. Even when higher-order corrections become important, the leading-order expression provides a useful interpretation of the role played by the degree fluctuations of the non-projected layer.

Perhaps the most surprising result emerging from the numerical analysis concerns the behaviour of the variance. Despite neglecting the heterogeneity of the non-projected layer, the Hypergeometric approximation reproduces the variance of the microcanonical ensemble remarkably well for all three empirical systems considered. At present we do not possess a complete theoretical explanation for this observation. Nevertheless, its robustness across networks with markedly different structural properties suggests that it is not accidental. Understanding why the Hypergeometric model captures the fluctuations of the co-occurrence distribution so accurately represents, in our opinion, an interesting theoretical problem deserving further investigation.

The present analysis also suggests a practical implication. Exact microcanonical validation based on Curveball randomization provides the most constrained statistical benchmark among the models considered but may become computationally prohibitive for very large bipartite systems. Analytical models therefore remain indispensable for many practical applications. Our results indicate that their use should not be guided solely by computational convenience but also by an understanding of how accurately they reproduce the first moments of the null distribution. In this respect, the empirical criterion introduced in this work provides a first indication of when the simplest analytical approximations are expected to recover the microcanonical benchmark with satisfactory accuracy.

The complementary strengths of the different analytical models naturally suggest the possibility of constructing improved approximations. As a proof of concept, we have explored a Hybrid Model ``CH'' combining the co-occurrence expectation predicted by the BiCM with the co-occurrence variance provided by the Hypergeometric approximation. Although the present analysis is limited to three empirical datasets, the encouraging results obtained suggest that improved analytical null models may be systematically constructed by independently optimizing different statistical properties of the null distribution. Assessing the general validity of this strategy, particularly for large bipartite networks where microcanonical validation is computationally unfeasible, represents an interesting direction for future research.

More generally, we believe that the perspective adopted here extends beyond the specific problem of bipartite network projections. In all statistical validation procedures based on randomized ensembles or maximum-entropy null models, the final inference depends on the probability distribution assigned to the test statistic under the null hypothesis. Different null models therefore represent different statistical assumptions rather than merely different randomization algorithms. Understanding how the constraints defining these null models propagate into the probability distribution of the test statistic is therefore essential for interpreting the outcome of hypothesis testing. From this viewpoint, the present work suggests that null models should be compared not only according to the structural constraints they preserve but also according to the statistical consequences that these constraints induce on the distribution of the quantities used for statistical inference.

\section*{Acknowledgments}

Work by R.N.M. is supported by \#NEXTGENERATIONEU (NGEU) and funded by the Ministry of University and Research (MUR), National Recovery and Resilience Plan (NRRP), project MNESYS (PE0000006) – A Multiscale integrated approach to the study of the nervous system in health and disease - DN. 1553 11.10.2022”

\end{document}